\documentclass[aps,floatfix]{revtex4-2}

\usepackage{graphicx,amsmath,bm}
\usepackage{epsfig}
\usepackage{amssymb}
\usepackage[colorlinks]{hyperref}
\usepackage[ansinew]{inputenc}
\usepackage{multirow}

\def\be{\begin{equation}}
\def\lan{\left\langle}
\def\ran{\right\rangle}
\def\ee{\end{equation}}
\def\barr{\begin{array}}
	\def\earr{\end{array}}

\def\dis{\displaystyle}
\def\ed{\end{document}}

\def\cs{{\bf s}}

\DeclareMathOperator{\erf}{erf}
\DeclareMathOperator{\erfc}{erfc}

\begin{document}

\title{Ordered Level Spacing Distribution in Embedded Random Matrix Ensembles}


\author{Priyanka Rao\textsuperscript{1}, H. N. Deota\textsuperscript{2} and N. D. Chavda\textsuperscript{1}}
\affiliation{\textsuperscript{1} Department of Applied Physics, Faculty of Technology and Engineering, The Maharaja Sayajirao University of Baroda, Vadodara-390001, India\\}
\affiliation{\textsuperscript{2} HVHP Institute of Post Graduate Studies and Research, Kadi Sarva Vishwavidyalaya, Gandhinagar, India}





\begin{abstract}
The probability distribution of the closest neighbor and farther neighbor spacings from a given level have been studied for interacting fermion/boson systems with and without spin degree of freedom constructed using an embedded GOE of one plus random two-body interactions. Our numerical results demonstrate a very good consistency with the recently derived analytical expressions using a $3 \times 3$ random matrix model and other related quantities by Srivastava et. al [{\it J. Phys. A: Math. Theor.} {\bf 52} 025101 (2019)]. This establishes conclusively that local level fluctuations generated by embedded ensembles (EE) follow the results of classical Gaussian ensembles.
\end{abstract}

\maketitle





\section{Introduction}
Random matrix theory (RMT) originally introduced by Wishart \cite{Wishart-1928} in statistics and further introduced by Wigner to study nuclear spectra \cite{Wigner-1955}, is now established as a good model to describe spectral fluctuations arising from complex quantum systems from a wide variety of fields like quantum chaos, finance \cite{Plerou-1999}, econophysics \cite{econophys-book}, quantum chromodynamics \cite{verbaarschot_QCD}, functional brain structures \cite{Seba-2003} and many more. These spectral fluctuations reveal whether the given complex quantum system is in regular(or integrable) or chaotic domain as well as they describe transition from regular to chaotic domain. One of the most popular measure of RMT widely used for this purpose is nearest neighbor spacing distribution (NNSD) $P(s)$, which tells us about the short-range correlations between nearest neighbors of energy-levels (or eigenvalues) of the complex quantum system. Dyson gave threefold classification of classical random matrix ensembles based on the symmetries present in their Hamiltonian viz. Gaussian Orthogonal Ensemble (GOE), Gaussian Unitary Ensemble (GUE) and Gaussian Symplectic ensemble (GSE) \cite{Haake}. For the case of GOE, which corresponds to quantum systems that are time reversal invariant without spin, the energy levels are correlated (corresponding to the chaotic behavior) and NNSD obeys the Wigner surmise which is essentially the GOE result $P(s) = (\pi/2) s \exp (-\pi s^2/4)$ \cite{Bohigas-1984}, whereas if the energy levels of a complex quantum system are uncorrelated (corresponding to the regular behavior), then the form of NNSD is given by the Poisson distribution $P(s) = \exp (-s)$ \cite{Berry-1977}. For a given set of energy levels, the construction of NNSD involves unfolding of the spectra to remove the variation in the density of eigenvalues \cite{Haake,Br-81}. Recently, NNSD has been used to study this transition from regular to chaotic domain in wormholes \cite{Gracia-2019} and open quantum systems \cite{Prosen2019}.

Complex systems can be represented in the form of a network and the spectral properties of these networks are now known to follow RMT. This opened a route to predict and control functional behavior of these complex systems \cite{Jalan-2015,Jalan-2018}. In some complex systems such as cancer networks, the short range correlations given by NNSD may give information only about the random connections in these networks. However, the long range correlations given by spectral rigidity can give further details about the underlying structural patterns in these systems \cite{Jalan-2014}. In such systems, study of measures giving long range correlations such as, the number variance and spectral rigidity  are important \cite{Br-81,Dyson1963,Berry1985,Berry1988}. These days, a very good alternative to NNSD called the ratio of level spacings \cite{Huse2007} is gaining a lot of attraction \cite{Jalan-2014,CK,CDK2014,NDC2015,Prosen2020,Relano2020} as it is simple to compute and no unfolding is needed as it is independent of the form of the density of the energy levels. The higher orders of ratio of spacings have also been studied in \cite{tekur2018-2,hospc2019}. The distribution intermediate of Poisson and GOE is described by Brody distribution \cite{Bordy1973}. Recently, intermediate semi-poissonian statistics \cite{Roy2017} and crossover random matrix ensembles \cite{sarkar2020} are also reported. Going beyond this, recently, the distribution of the closest neighbor (CN) spacing, $s_{CN}$ and farther neighbor (FN) spacing, $s_{FN}$ from a given level are introduced \cite{sashi-2019}. The distribution of $s_{CN}$ spacings is important in the context of perturbation theory, as the contribution from the CN is prominent due to smaller energy spacing \cite{purt-1}. The distribution of $s_{FN}$ spacings is complementary to that of $s_{CN}$. It is important to note that the ratio of two consecutive level spacings introduced in \cite{Huse2007} is given by $\tilde{r}=\frac{s_{CN}}{s_{FN}}$. The numerical results for the integrable circle billiard, fully chaotic cardioid billiard, standard map with chaotic dynamics and broken time reversal symmetry, and the zeros of the Riemann zeta function are shown to be in very good agreement with the analytical formulas derived in \cite{sashi-2019} for the random matrix ensembles GOE, GUE,
and GSE based on a $3\times3$ matrix modeling and Poisson spectra. In the present work, we analyze distributions of $s_{CN}$ and $s_{FN}$ using random matrix ensembles defined by one- plus chaos generating two-body interactions operating in
many particle spaces, to conclusively establish that these measures are universal and local level fluctuations generated by many-particle interacting systems follow the results of classical Gaussian ensembles \cite{vkbk01,Weiden,Gomez}. These ensembles are generically called embedded ensembles
of (1+2)-body interactions or simply EE(1+2) and their GOE random matrix version is called EGOE(1+2). These models, both for fermion and boson systems, including spin degree of freedom and without spin, have their origin in nuclear
shell model and the interacting boson model \cite{KC2018}.

Now, it is very well established that EGOE(1+2) ensembles apply in a generic way to isolated finite interacting many-particle quantum systems such as nuclei, atoms, quantum dots, small metallic grains, interacting spin systems modeling quantum computing core and so on  \cite{Weiden,Kota-book}. For sufficiently strong interaction, EGOEs exhibit average-fluctuation separation in eigenvalues with the smoothed eigenvalue density being a corrected Gaussian and the local fluctuations are of GOE type \cite{Br-81,Gomez,Lec-08}. Recently these models have also been used successfully in understanding high energy physics related problems. Random matrix models with two-body interactions [EGOE(2)] among complex fermions are known as complex Sachdev-Ye-Kitaev models in this area \cite{Davison2017, Bulycheva2017,Rosenhaus2019}. EGOE(1+2) can be defined for fermions and bosons with spin degree of freedom and also with many other symmetries \cite{Gomez,Kota-book}. Now we will give a preview.

The rest of this paper is organized as follows. In Section \ref{sec:2}, we briefly describe the construction of five different EGOEs used in the present paper. Analytical results for the probability distribution for the CN spacings and the FN spacings are discussed in section \ref{sec:3}. Section \ref{sec:4} presents the numerical results for the probability distribution for the CN spacings and the FN spacings. Finally, we draw conclusions in Section \ref{sec:5}.


\section{Embedded ensembles for fermion and boson systems}
\label{sec:2}
In this section, we describe the construction of various embedded random matrix ensembles (EE) used in this paper. Let us begin with EE for spinless systems. For defining such ensembles, one can consider a system of $m$ spinless particles (fermions or bosons) which are to be distributed in $N$ single particle (sp) states and interacting via (1+2) body interaction. Let these $N$ sp states be denoted by $|v_{i} \rangle$ where $i=1,2,3,...,N$. A two-particle Hamiltonian matrix can be constructed and then it can be further embedded to the $m$-particle space by using the concepts of direct product space and Lie algebra. With GOE embedding these ensembles are called EGOE(2) [or BEGOE(2) for bosons].
For such a system, one can define the two body Hamiltonian matrix by the expression :
\begin{equation}
V(2)= \sum_{\alpha,\gamma} V_{2;\alpha,\gamma} \ A_{2,\alpha}^{\dag} A_{2,\gamma}
\label{2bodyH}
\end{equation}
where the term $V_{2;\alpha,\gamma}$ in Eq.\eqref{2bodyH} is the Gaussian random variate with zero mean and constant variance,
\begin{equation}
\overline{V_{2;\alpha,\gamma} V_{2;\alpha',\gamma'}} = \nu_{0}^{2} (1+\delta_{\alpha,\alpha',\gamma,\gamma'})
\label{2bodyH_1}
\end{equation}
where the overbar denotes the ensemble average and $\nu_{0}=1$ without the loss of generality.
For fermions, $A_{2,\alpha}^{\dag} = a_{v_{1}}^{\dag} a_{v_{2}}^{\dag}$ ; $A_{2,\alpha} = (A_{2,\alpha}^{\dag})^{\dag}$ $( v_{1} < v_{2} )$.
Whereas for bosons, $A_{2,\alpha}^{\dag} = C \ b_{v_{1}}^{\dag} b_{v_{2}}^{\dag}$;  $A_{2,\alpha} = (A_{2,\alpha}^{\dag})^{\dag}$ $( v_{1}  \leq v_{2} )$, where, $C$ is the normalization constant given by $C = \Pi_{i=1}^{2} (v_{i} !)^{1/2}$ and $\alpha$ simplifies the notation of indices.
Also, $a^{\dag}_{v_{i}}$ and $a_{v_{i}}$ are the creation and annihilation operators respectively for fermions and $b^{\dag}_{v_{i}}$ and $b_{v_{i}}$ are the creation and annihilation operators respectively for bosons.
One should also note that the dimension of Hamiltonian matrix for fermions would be $d(N,m)=\binom{N}{m}$ and for bosons $d(N,m) = \binom{N+m-1}{m}$, with the two-body independent matrix elements (TBME) being $\frac{d(N,2)(d(N,2)+1)}{2}$ for both. $V_{2;\alpha,\gamma}$ in Eq.\eqref{2bodyH}, are anti-symmetrized TBME for fermions and symmetrized TBME for bosons.

In such a manner, one can construct an embedded two-body random matrix ensemble. When the mean field one body part is added to the Hamiltonian, they are generally called as one plus two-body random matrix ensembles [EGOE(1+2)].
Thus, with random two-body interactions $V(2)$, we can define the Hamiltonian of EGOE(1+2) as follows,
\begin{equation}
H = h(1) + \lambda \{V(2)\}.
\label{eq.egoe1}
\end{equation}
Here, $h(1) = \sum_i \epsilon_i n_i$ is the one-body part of the Hamiltonian. The sp energies are defined as $\epsilon_i$ and $n_i$ are number operators acting on sp states. $\lambda$ is the two-body interaction strength and notation $\{\;\}$ denotes an ensemble. The $V(2)$ matrix is chosen to be a GOE in two-particle spaces \cite{Kota-book}. Due to ($1+2$)-body nature of the interaction, many of the matrix elements of $H(m)$ for $m > 2$ are zero and the nonzero
matrix elements are linear combinations of the sp energies and the TBMEs.

Going beyond spin-less systems, we have considered three embedded ensembles with spin degree of freedom. For fermions with spin $\cs=1/2$ degree of freedom, we have EGOE(1+2)-$\cs$ \cite{ksc-2006}. Here, the interaction $V(2)$ will have two parts as the two-particle spin $s = 0$ and 1 giving EGOE(1+2)-$\cs$ Hamiltonian $H=h(1)+\lambda_0 \{V^{s=0}(2)\}+\lambda_1 \{V^{s=1}(2)\}$. For bosons with spin degree of freedom, we have considered the following two embedded ensembles: (i) for two species boson systems with a fictitious ($F$) spin $1/2$ degree of freedom, we have BEGOE(1+2)-$F$ \cite{vyas-12}. Here also, the interaction $V(2)$ will
have two parts as the two-particle $F$-spin $f = 0$ and 1 giving  BEGOE(1+2)-$F$ Hamiltonian $H =h(1)+\lambda_0 \{V^{f=0}(2)\}+\lambda_1 \{V^{f=1}(2)\}$. (ii) for bosons with spin one degree of freedom, we have BEGOE(1+2)-$S1$ \cite{Deota}. Here, the interaction $V(2)$ will have three parts as the two-particle spins are $s = 0$, 1 and 2 giving  BEGOE(1+2)-$S1$ Hamiltonian $H =h(1)+\lambda_0 \{V^{s=0}(2)\}+\lambda_1 \{V^{s=1}(2)\} + \lambda_2 \{V^{s=2}(2)\}$. Note that, the sp levels ($\Omega$) defining one-body part $h(1)$ for embedded ensembles will have $(2s+1)$ degeneracy. For EGOE(1+2)-$\cs$ and BEGOE(1+2)-$F$, the sp levels  will be doubly degenerate ($N=2\Omega$), while for BEGOE(1+2)-$S1$, they will be triply degenerate ($N=3\Omega$). In all the five ensembles, without loss of generality, we choose the average spacing between the sp levels to be unity so that all strength of interactions are unit-less.

\section{Ordered level spacing distribution}
\label{sec:3}
Let us consider an ordered set of unfolded eigenvalues (energy levels) $E_n$, where
$n=1,2,...,d$. The Nearest-Neighbor Spacing is given by $s_n = E_{n+1} -
E_{n}$. Then, the closest neighbor (CN) spacing is defined as $s_n^{CN}=\min\{s_{n+1},s_n\}$ and the farther neighbor (FN) spacing is defined as $s_n^{FN}=\max\{s_{n+1},s_n\}$. The probability distribution for the closest neighbor (CN) spacings is denoted by
$P_{CN}(s)$ and for the farther neighbor (FN) spacings is denoted by
$P_{FN}(s)$. If the system is in integrable domain, NNSD is Poisson. Then $P_{CN}(s)$ and $P_{FN}(s)$ are given by,
\begin{equation}
P_{CN}^{P}(s)=\dis 2 \exp(-2s)\;
\label{eq1}
\end{equation}
and
\begin{equation}
P_{FN}^{P}(s)=\dis 2 \exp(-s) [1-\exp(-s)]\;,
\label{eq2}
\end{equation}
respectively. Similarly, if the system is in chaotic domain, NNSD is GOE and is derived using $3 \times 3$ real symmetric matrices.
Then $P_{CN}(s)$ and $P_{FN}(s)$ are given by \cite{sashi-2019},

\begin{equation}
\begin{array}{lcl}
P_{CN}^{GOE}(s)&=&\frac{a}{\pi} s \;\exp(-2 a s^2) [ 3\sqrt{6 \pi a}\;s - \pi \exp(\frac{3a}{2} s^2) \\ &&\times (a s^2-3) \erfc(\sqrt{\frac{3a}{2}}\;s) ]\;
\label{pcngoe}
\end{array}
\end{equation}
and
\begin{equation}
\begin{array}{lcl}
P_{FN}^{GOE}(s)&=& \frac{a}{\pi} s\; \exp(-2 a s^2)\; [\pi \exp(\frac{3a}{2} s^2) \\ && \times (a s^2-3)\; \{\erf(\sqrt{\frac{a}{6}}\;s)- \erf(\sqrt{\frac{3a}{2}}\;s)\} \\ && + \sqrt{6 \pi a}\;s (\exp(\frac{4a}{3}\;s^2)-3) ] \;
\label{pfngoe}
\end{array}
\end{equation}
respectively. Here $a=\frac{27}{8\pi}$. It is important to note that $2P(s)=P_{CN}(s)+P_{FN}(s)$.
For small spacings $s$, $P_{CN}^{GOE}(s)$ shows level-repulsion similar to the NNSD and $P_{FN}^{GOE}(s) \propto s^4$. While for large $s$, $P_{FN}^{GOE}(s) \propto \exp(\frac{-2a}{3}s^2)$.  For GOE, the average value $\lan s_{CN} \ran=\frac{2}{3}$ and for Poisson it is $\frac{1}{2}$. However, the average value $\lan s_{FN} \ran=\frac{4}{3}$ for GOE and $\frac{3}{2}$ for Poisson.

Here, spectral fluctuations in EEs for fermion and boson systems with
and without spin degree of freedom are studied using $P_{CN}$ and $P_{FN}$ and it is found that these forms of the distributions are universal.
Let us add that the ensembles without spin and with spin degree of freedom are used to represent the quantum many particle systems with interactions \cite{Kota-book}. We present the numerical results in the next section.

\section{Numerical results}
\label{sec:4}

In order to study closest neighbor spacing distribution $P_{CN}(s)$ and farther neighbor spacing distribution $P_{FN}(s)$, we consider the following five EGOEs
in many particle spaces:

\begin{enumerate}
	
	\item {EGOE(1+2) for $m=6$ fermions in $N=12$ sp states with $H$ matrix of
		dimension 924. The interaction strength $\lambda = 0.1$. See
		Ref. \cite{vkbk01} for details.}
	
	\item {BEGOE(1+2) for $m=10$ bosons in $N = 5$ sp states with $H$ matrix of
		dimension 1001. The interaction strength $\lambda = 0.06$. See
		Ref.\cite{cpk-2003,ckp-2012} for details.}

	\item {EGOE(1+2)-$\cs$ for $m=6$ fermions occupying $\Omega=8$ sp levels (each
		doubly degenerate) with total spin $S=0$ and $S=1$ giving the $H$ matrices of
		dimensions 1176 and 1512 respectively. The interaction strength
		$\lambda =\lambda_0 =\lambda_1 = 0.1$. See Ref. \cite{ksc-2006, mkc-2010}for
		details.}
	
	\item {BEGOE(1+2)-$F$ for $m=10$ bosons occupying $\Omega = 4$ sp levels (each
		doubly degenerate) with total $F$-spin $F=0, 2$ and $F=F_{max}=5$ giving the $H$
		matrices of dimensions 196, 750 and 286. The interaction strength
		$\lambda =\lambda_0 =\lambda_1= 0.08$. See Ref.\cite{vyas-12,KC-17} for details.}
	
	\item {BEGOE(1+2)-$S1$ for $m=8$ bosons occupying $\Omega = 4$ sp levels (each
		triply degenerate) with total spin $S=4$ giving the $H$ matrix of dimension
		1841. The interaction strength $\lambda = \lambda_0 =\lambda_1
		= \lambda_2= 0.2$; see Ref.\cite{Deota} for details.}
	
\end{enumerate}

In the present analysis, an ensemble of 500 members is used for all the examples. The single-particle
energies defining $h(1)$ are chosen as $\epsilon_i = (i + 1/i)$. It is important to note that as $\lambda$ increases in these embedded ensembles, (both fermion and boson), there is Poisson to GOE transition in level fluctuations at $\lambda = \lambda_C$ and Breit-Wigner to Gaussian transition in strength functions (also known as local density of states) at $\lambda =  \lambda_F > \lambda_C$. Also, they generate a third chaos marker at $\lambda =  \lambda_t > \lambda_F$, a point or a region where thermalization occurs. The values of $\lambda$ in the ensemble calculations are chosen sufficiently large so that there is enough mixing among the basis states and the system is in the Gaussian domain, i.e. $\lambda > \lambda_F$. For EGOE(1+2) \cite{vkbk01} and EGOE(1+2)-$\cs$ \cite{ksc-2006,mkc-2010},  fermion systems are always in Gaussian domain with $\lambda=0.1$. For spin-less boson BEGOE(1+2), $\lambda=0.06$ is sufficiently large so that the system is in Gaussian domain \cite{cpk-2003,ckp-2012}. Similarly, for boson ensembles with spin degree, BEGOE(1+2)-$F$ with $\lambda=0.08$ \cite{vyas-12,KC-17} and BEGOE(1+2)-$S1$ with $\lambda=0.2$ \cite{Deota}, again the systems exhibit GOE level fluctuations and the eigenvalue density as well as strength functions are close to Gaussian.

\begin{figure}[!tbh]
	\centering
\begin{tabular}{cc}
		\includegraphics[width=0.4\textwidth]{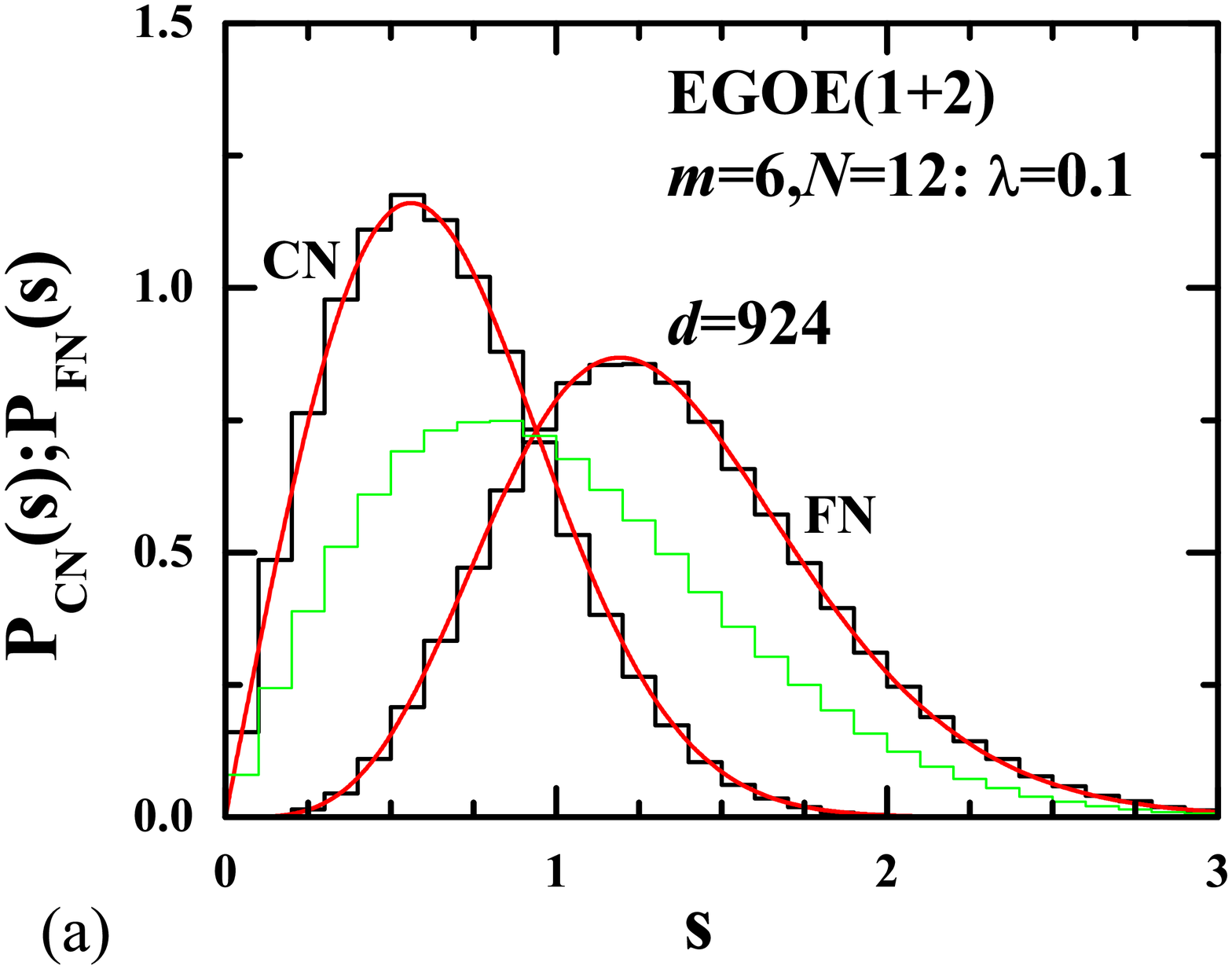}&
		\includegraphics[width=0.4\textwidth]{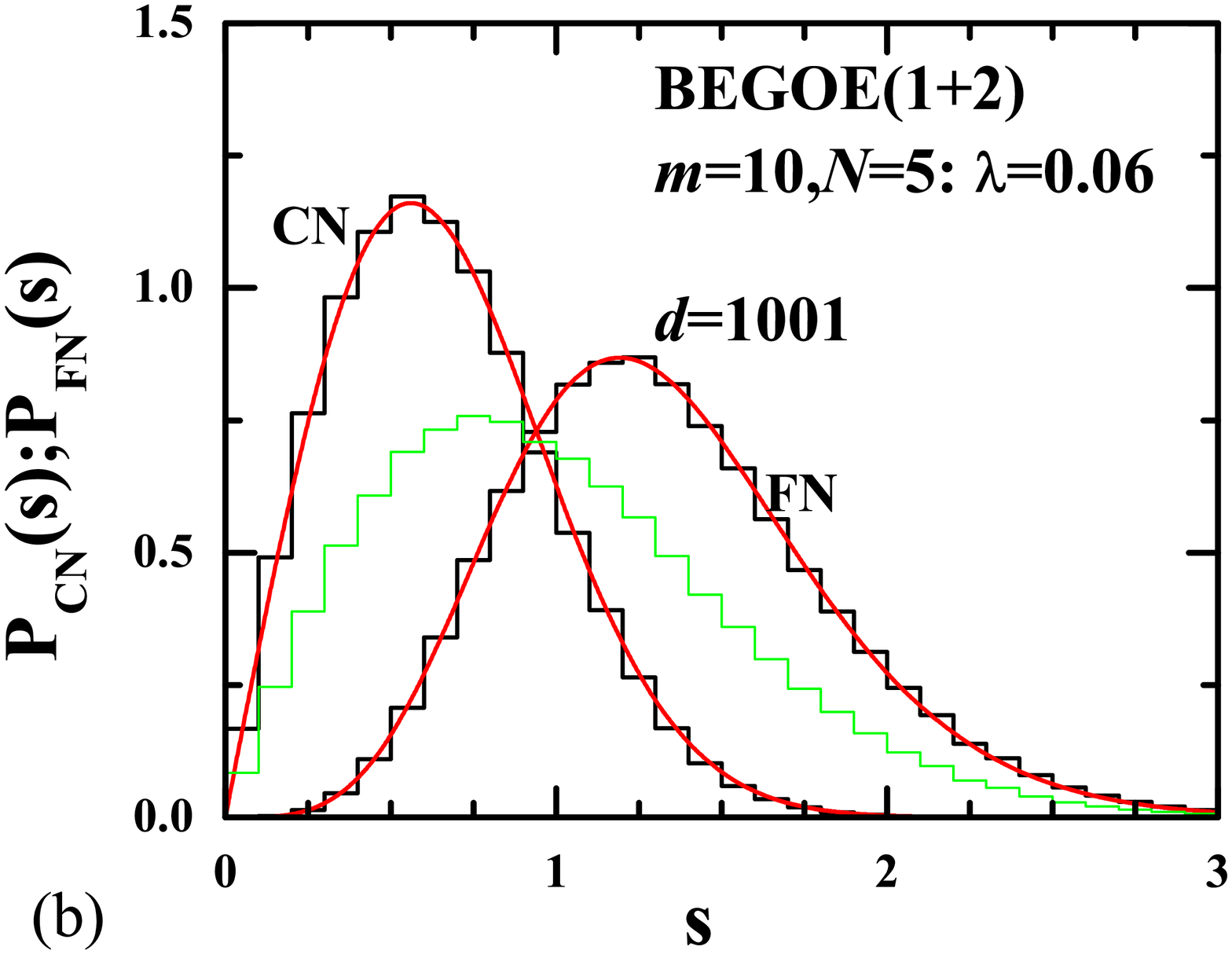}
\end{tabular}
	\caption{The closest neighbor spacing distribution $P_{CN}(s)$ and farther neighbor spacing distribution $P_{FN}(s)$ (black histograms) for a 500 member (a) EGOE(1+2) ensemble and (b) BEGOE(1+2) ensemble. The red smooth curves are due to corresponding Eqs. (\ref{pcngoe}) and (\ref{pfngoe}). The NNSD is shown by green histogram for comparison.}\label{egoe}
\end{figure}

\begin{figure}[!tbh]
	\centering
	\includegraphics[width=0.7\textwidth,bb=95 0 630 620]{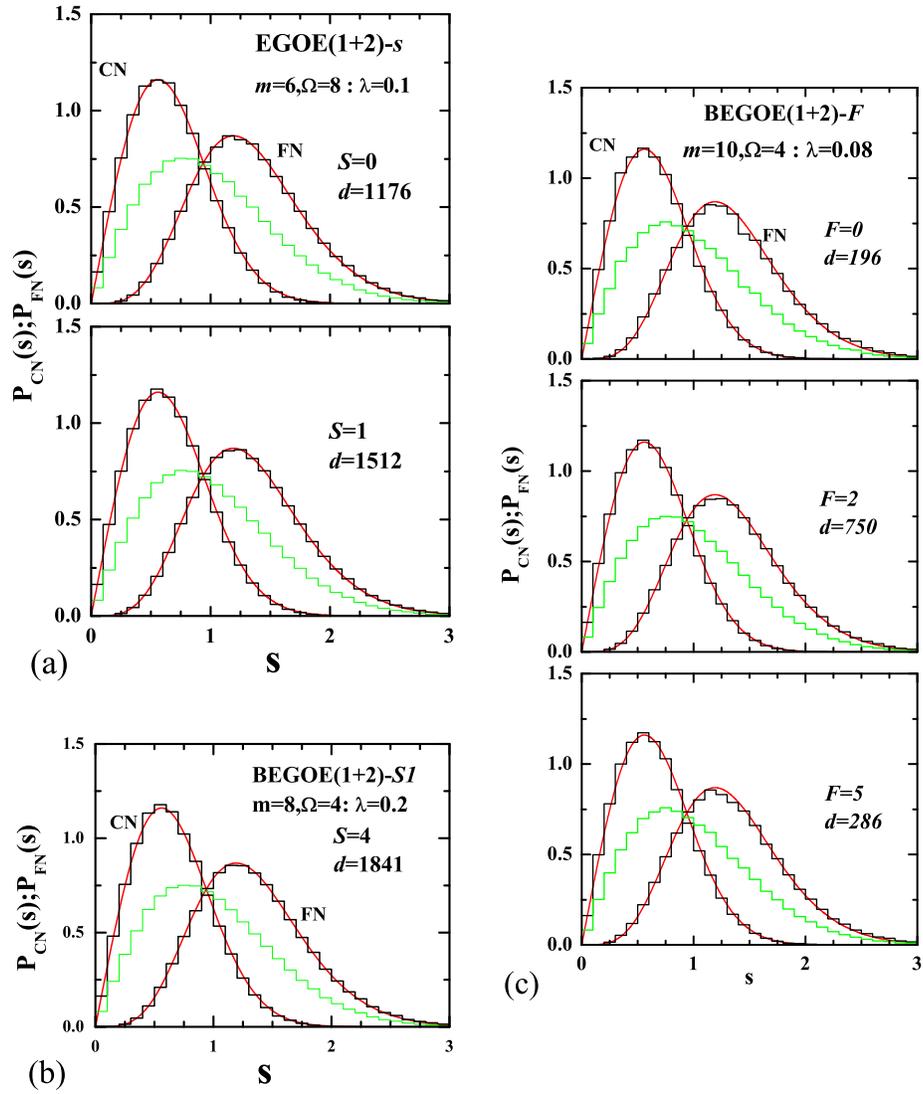}
	\caption{The closest neighbor spacing distribution $P_{CN}(s)$ and farther
		neighbor spacing distribution $P_{FN}(s)$ for (a) EGOE(1+2)-$\cs$ ensemble for spin values $S=0$ and $1$ (b) BEGOE(1+2)-$S1$
		ensemble for spin value $S=4$ and (c) BEGOE(1+2)-$F$
		ensemble for spin values $F=0,2$ and $5$. See Figure \ref{egoe} and
		text for details.}\label{egoe-s}
\end{figure}

\begin{figure}[!tbh]
	\centering
	\includegraphics[width=0.4\textwidth]{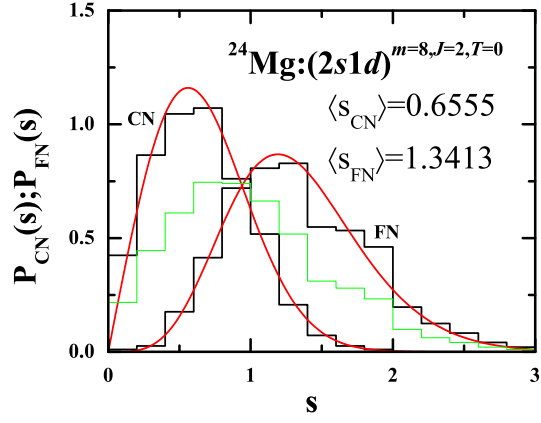}
	\caption{The closest neighbor spacing distribution $P_{CN}(s)$ and farther
		neighbor spacing distribution $P_{FN}(s)$ vs. $s$ for Nuclear shell model example:
		$^{24}$Mg with 8 nucleons in the ($2s1d$) shell with angular momentum
		$J = 2$ and isospin $T = 0$. The matrix dimension is
		1206 and all levels are used in the analysis. See Ref.\cite{Lec-08} for
		further details. The skewness and excess parameters are $\gamma_1 = 0.139$ and $\gamma_2 =-0.061$. $\lan s_{CN} \ran$ and $\lan s_{FN} \ran$ values are also given in the figure.}\label{fig:sm}
\end{figure}

In the analysis, $P_{CN}(s)$ and  $P_{FN}(s)$ are obtained using the following procedure. First the spectrum for each member of the ensemble is unfolded using the procedure described in \cite{Lec-08}, with the smooth density as a corrected Gaussian with corrections involving up to 4-6th order moments of the density function so that the average spacing is unity. The ensemble averaged skewness ($\gamma_1$) and excess ($\gamma_2$) parameters are shown in Table \ref{table1} for all the examples of embedded ensembles analyzed in the present work. The histograms for $P_{CN}(s)$ and $P_{FN}(s)$ are constructed using the central 80\% part of the spectrum with the bin size equal to 0.1. The results for embedded ensembles without spin, EGOE(1+2) and BEGOE(1+2), are shown in Figure \ref{egoe}. Similarly, the results for EE with spin degree of freedom, EGOE(1+2)-$\cs$ and BEGOE(1+2)-$F$ and BEGOE(1+2)-$S1$, are shown in Figure \ref{egoe-s}. A very good agreement is observed between the numerical embedded ensemble results and the theoretical predictions given by Eqs.\eqref{pcngoe} and \eqref{pfngoe} for all the examples. The ensemble averaged values of $\langle s_{CN} \rangle$ and  $\langle s_{FN} \rangle$, for all the examples, are given in Table \ref{table2}. They are found to be very close to corresponding GOE estimates. In addition to this, we have also analyzed shell model example which is a typical member of EGOE(1+2)-$JT$ \cite{vkbk01}. This ensemble is usually called TBRE in literature \cite{papen-2007}. The result is shown in Figure \ref{fig:sm}. Here also the shell model results along with the calculated averages are consistent with the theoretical predictions.

\begin{figure}[tbh!]
	\centering
	\includegraphics[width=0.4\textwidth]{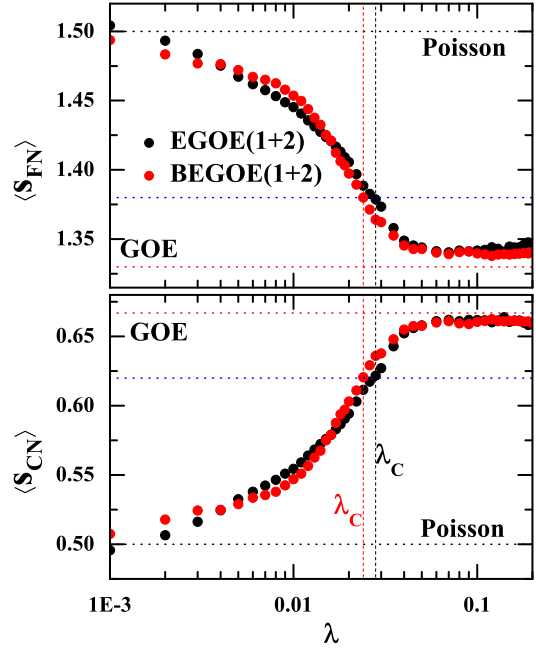}
	\caption{Ensemble averaged values of $\lan s_{CN} \ran$ (lower panel) and $\lan s_{FN} \ran$ (upper panel) as a function of the two-body strength of interaction $\lambda$, obtained for EGOE(1+2) ensemble with $(m,N)=(6,12)$ (black circles) and BEGOE(1+2) ensemble with
		$(m,N)=(10,5)$ (red circles). In the calculations sp energies are drawn from the center of a GOE. The vertical dash-lines represent the position of $\lambda_C$ for the corresponding EGOE(1+2) and BEGOE(1+2) examples. In each calculation, an ensemble of 500 members is used. The horizontal dotted-lines represent Poisson estimate (black), GOE estimate (red) and ${\lan s_{CN} \ran}_C=0.62$ (and ${\lan s_{FN} \ran}_C=1.38$). See text for further details.}\label{poi-to-G}
\end{figure}

Going further, it is also possible to study a transition from Poisson to GOE in terms of $\langle s_{CN} \rangle$ and $\langle s_{FN} \rangle$ for EGOE(1+2) and  BEGOE(1+2) ensembles as these ensembles demonstrate Poisson to GOE transition in level fluctuations with increase in the strength of the two-body interaction $\lambda$  \cite{NDC2015,vkbk01,vyas-12,cpk-2003,mkc-2010}. We have computed $\langle s_{CN} \rangle$ and $\langle s_{FN} \rangle$ for spin-less fermion
and boson ensembles by varying the interaction strength $\lambda$. Figure \ref{poi-to-G} represents these results. It is clearly seen that for lower values of $\lambda$, the values of $\langle s_{CN} \rangle$ and $\langle s_{FN} \rangle$ are close to Poisson, which gradually reach the GOE value with increase in $\lambda$. Therefore, there is a transition from Poisson to GOE form in $P_{CN}(s)$ (and also in $P_{FN}(s)$). With this it is possible to define a chaos marker $\lambda_C$ such that for $\lambda > \lambda_C$, the level fluctuations follow GOE. This transition occurs when the interaction strength $\lambda$ is of the order of the spacing $\Delta$ between the states that are directly coupled by the two-body interaction. 
In the past, the NNSD \cite{KS1999} and the distribution of ratio
of consecutive level spacings \cite{CDK2014} have been used to study Poisson-to-GOE transition by constructing suitable random matrix model and the transition parameters were used to identify the chaos marker $\lambda_C$ in the embedded ensembles \cite{NDC2015,vkbk01,vyas-12,cpk-2003,mkc-2010}. Corresponding to the critical values of these transition parameters required for onset of GOE fluctuations, we found the critical value of $\langle s_{CN} \rangle$, ${\langle s_{CN} \rangle}_C=0.62$ (and ${\langle s_{FN} \rangle}_C=1.38$). This is represented by blue dotted lines in figure \ref{poi-to-G}. ${\langle s_{CN} \rangle}_C =0.62$ gives $\lambda_C \simeq 0.028$ for EGOE(1+2) example and $\lambda_C \simeq 0.024$ for BEGOE(1+2) example. These values are shown by dashed vertical lines in figure \ref{poi-to-G} and are close to the previously obtained results \cite{NDC2015,Kota-book}. Therefore, these measures can also be utilized to identify $\lambda_C$ marker using $P_{CN}(s)$.
In the past, the criterion for the chaos marker $\lambda_C$ for EGOE(1+2) models \cite{Kota-book,cpk-2003}, based on the perturbation theory was derived by Jacquod and Shepelyansky \cite{Jac97}. The validity of the perturbation theory gives $\lambda_C$. Hence, it is important to analyze $P_{CN}(s)$ distribution and related measures in the context of onset of chaos in embedded ensembles. This is for future.

\section{Conclusion}
\label{sec:5}
In this paper, we have studied the closest neighbor spacing distribution $P_{CN}(s)$ and the farther neighbor spacing distribution $P_{FN}(s)$ for interacting fermion/boson systems with and without spin degree of freedom. The system Hamiltonian is modeled by an embedded GOE of one plus two-body interactions[EGOE(1+2)]. In the past it was shown \cite{Br-81} that only with proper spectral unfolding, EEs exhibit GOE level fluctuations. Our numerical results for various examples of fermion/boson system and shell model, are consistent with the recently derived analytical expressions using a $3 \times 3$ random matrix model and other related quantities\cite{sashi-2019}. This establishes that these analytical expressions are universal. Also, it shows that for strong enough interaction, the local level fluctuations generated by EEs follow the results of classical Gaussian ensembles.

\section*{Acknowledgement}
One of the authors (NDC) thanks V K B Kota for useful discussions. PR and NDC acknowledge financial support from Science and Engineering Research Board, Department of Science and Technology(DST), Government of India [Project No.: EMR/2016/001327].


\begin{table}[!tbh]
	\centering
	\caption{The ensemble averaged skewness $\gamma_1$ and excess $\gamma_2$ parameters for various EE examples used.}
	
	\begin{tabular}{|c|c|c|}
		\hline
		EE & $\gamma_1$ & $\gamma_2$\\
		\hline
		EGOE(1+2) & 0.0008 & -0.3431\\
		&&\\
		BEGOE(1+2) & 0.0922 & -0.2329\\
		&&\\
		EGOE(1+2)-$\cs$   & & \\
		$S=0$ & 0.0202 & -0.3034\\
		$S=1$ & 0.0178 & -0.3352\\
		&&\\
		BEGOE(1+2)-$F$    & & \\
		$F=0$ & 0.0088 & -0.3114\\
		$F=2$ & 0.0469 & -0.3129\\
		$F=5$ & 0.0677 & -0.2569\\
		&&\\
		BEGOE(1+2)-$S1$   & & \\
		$S=4$ & 0.0349 & -0.1111\\
		\hline
	\end{tabular}\label{table1}
\end{table}

\begin{table}[!tbh]
	\centering
	\caption{Average values of the closest neighbor ($\lan s_{CN} \ran$) and farther neighbor ($\lan s_{FN} \ran$) spacings obtained numerically for various EE examples used in the present paper. Average values obtained from theory for Poisson and GOE are also given.}\label{table2}
	\begin{tabular}{|c|c|c|}
		\hline
		EE & $\lan s_{CN} \ran$ & $\lan s_{FN} \ran $\\
		\hline
		EGOE(1+2) & 0.6613 & 1.3417\\
		&&\\
		BEGOE(1+2) &0.6600 & 1.3401\\
		&&\\
		EGOE(1+2)-$\cs$   & & \\
		$S=0$ & 0.6616 & 1.3411\\
		$S=1$ & 0.6625 & 1.3409\\
		&&\\
		BEGOE(1+2)-$F$    & & \\
		$F=0$ &0.6585 & 1.3421\\
		$F=2$ &0.6600 & 1.3404\\
		$F=5$ &0.6578 & 1.3420\\
		&&\\
		BEGOE(1+2)-$S1$   & & \\
		$S=4$ &0.6600 & 1.3401\\
		&&\\
		Poisson &  $\frac{1}{2}$ & $\frac{3}{2}$ \\
		&&\\
		GOE & $\frac{2}{3}$ &  $\frac{4}{3}$ \\
		\hline
	\end{tabular}
\end{table}


\end{document}